\documentclass[aps,prl,twocolumn,superscriptaddress,showpacs,amsmath,amssymb,floatfix,english]{revtex4}

\usepackage{graphicx}
\usepackage{dcolumn}
\usepackage{bm}
\usepackage{xspace, epstopdf}

\usepackage[T1]{fontenc}
\usepackage[latin1]{inputenc}
\usepackage{float}
\usepackage{babel}

\newcommand{\test}{QPC\xspace}
\newcommand{\Go}{$G_0$\xspace}

\newcommand{\G}{conductance\xspace}

\newcommand{\po }{pinch-off\xspace}

\begin{document}

\title{Charge rearrangement and screening in a quantum point contact}

\author{S. L\"uscher}
\affiliation{Department of Physics, Stanford University, Stanford,
California 94305}
\author{L. S. Moore}
\affiliation{Department of Physics, Stanford University, Stanford,
California 94305}
\author{T. Rejec}
\affiliation{Department of Physics, Ben Gurion University of the
Negev, Beersheva, Israel}
\author{Yigal Meir}
\affiliation{Department of Physics, Ben Gurion University of the
Negev, Beersheva, Israel}
\author{Hadas Shtrikman}
\affiliation{Submicron Center, Weizmann Institute of Science,
Rehovot, Israel}
\author{D. Goldhaber-Gordon}
\affiliation{Department of Physics, Stanford University, Stanford,
California 94305}

\date{\today}

\begin{abstract}
Compressibility measurements are performed on a quantum point
contact (QPC). Screening due to mobile charges in the QPC is
measured quantitatively, using a second point contact. These
measurements are performed from pinch-off through the opening of the
first few modes in the QPC. While the measured signal closely
matches a Thomas-Fermi-Poisson prediction, deviations from the
classical behavior are apparent near the openings of the different
modes. Density functional calculations attribute the deviations to a
combination of a diverging density of states at the opening of each
one-dimensional mode and exchange interaction, which is strongest
for the first mode.
\end{abstract}


\maketitle The simplest of mesoscopic systems is a point contact, a
narrow constriction between two electron reservoirs. Conductance
measurements through such a quantum point contact  (QPC) reveal
steps in units of $G_0 \equiv 2e^2/h$. The physics of these steps is
well understood: for a QPC adiabatically connected to the
reservoirs, the transmission coefficient of each mode is either zero
or one. The number of such modes increases, as the width of the QPC
is increased, leading according to the Landauer formula to a series
of quantized steps in conductance.
 Such conductance measurements, however, also reveal our limited
knowledge of the physics of QPCs at low densities. As a QPC is just
being opened up, its conductance pauses around $0.7 \times G_0$,
before rising to the first full-channel plateau. This ``0.7
structure'' has been one of the prime puzzles in mesoscopic physics
\cite{Thomas96}. Longer 1D wires show a similar, and likely related,
structure at $0.5 \times G_0$ \cite{dePicciotto}. These features
have been variously attributed to spontaneous spin polarization in
the QPC \cite{Wang, Spivak, Bruus}, to Luttinger liquid behavior
\cite{Matveev}, or to the formation of a localized moment at the
QPC \cite{Wingreen, Hirose, Sara}, together with the resulting Kondo
effect.

To shed more light on the physics of QPCs near \po, we report here
measurements of the compressibility of electrons in the channel of
the QPC. To measure compressibility, one applies a potential to an
electrode on one side of a structure of interest, and measures the
potential on the other side. Compressibility measurements have
helped elucidate how charge carriers arrange themselves in two
dimensions when kinetic energy is dominated by interactions, whether
in the quantum Hall regime \cite{Eisenstein} or at low density and
high effective mass \cite{Sivan}. Augmented by local electrostatic
detectors, such measurements have produced striking images of how
individual carriers localize in these same regimes \cite{Yacoby}.
Compressibility measurements using nanofabricated electrostatic
detectors such as QPCs have also become a standard tool to probe
transitions between charge states of a quantum dot or two coupled
quantum dots~\cite{Field, Sprinzak, Elzerman}. In this letter we
present analogous measurements, using a QPC as an electrostatic
detector but measuring in this case the charge configuration of a
second QPC instead of a quantum dot, c.f. Ref. \cite{Castleton}.

Compressibility of perfect 1D systems is a textbook problem. The
diverging density of states should give strong compressibility at
low carrier density. In a multimode 1D system, one would expect that
a similar signature should occur at the opening of each mode. In
contrast, both experiment and numerical simulation of our short 1D
wire show that the enhancement in compressibility at the opening of
the first mode is significantly different from that at the opening
of higher modes. In retrospect, this might have been expected
because at the opening of the first mode the total density is small
and thus exchange effects are also important. In addition, the
distinctive ``0.7'' transport feature might have had a counterpart
in compressibility. As it turns out, the simulated compressibility
feature associated with the 0.7 regime was too weak to observe
experimentally with our current sensitivity.

The devices in this experiment were fabricated on a GaAs/AlGaAs
heterostructure, containing a two-dimensional electron gas (2DEG)
$70$~nm below the surface. The
electron density $n_s=2 \cdot 10^{11}$~cm$^{-2}$ and mobility
$\mu=2.3 \cdot 10^6$~${\rm cm}^2/{(\rm Vs)}$ at 4.2~K. A schematic
of the measured devices is shown in the inset of Fig.~1(a). Two
QPCs, with lithographic widths of 330~nm and 350~nm, respectively,
are separated by an 80~nm wide gate. Although the device is
symmetric in design, the two QPCs play fundamentally different roles
in our experiment: the lefthand QPC serves as a detector, sensitive
to charge rearrangements in the righthand QPC. For clarity, we will
henceforth refer to the righthand QPC as ``the QPC'' and the
lefthand QPC as ``the detector.'' Our data support the assumption
that the primary interaction between the two QPCs is electrostatic.
Two nominally-identical devices, each containing a QPC and a
detector, were measured in a $^3$He cryostat with 300~mK base
temperature. The extensive measurements taken on one QPC-detector
pair are presented in this paper; consistent behavior was also found
after thermal cycling to room temperature. Measurements of another
nominally-identical QPC-detector pair were used to confirm
qualitative features of the data presented here.

Fig.~1(a) and (b) show the linear and non-linear differential
conductance, respectively, of a QPC with the classic signatures of
0.7 structure.  The charge detector signal (Fig.~1(c)) is measured
as the QPC is opened from \po through the third plateau. The full
range of $V_{\rm qpc}$ is broken into 14 shorter measurements and
the detector readjusted for each, to keep the detector \G in its
sensitive, near-linear regime. Each of the 14 traces in Fig.~1(c)
covers a 100~mV range in $V_{\rm qpc}$, with consecutive traces
overlapping by 50~mV. To reduce noise we averaged each trace over 40
measurements.  No striking features are visible in these raw data.

\begin{figure}[h!]
  \begin{center}
  \includegraphics[width=0.5 \textwidth]{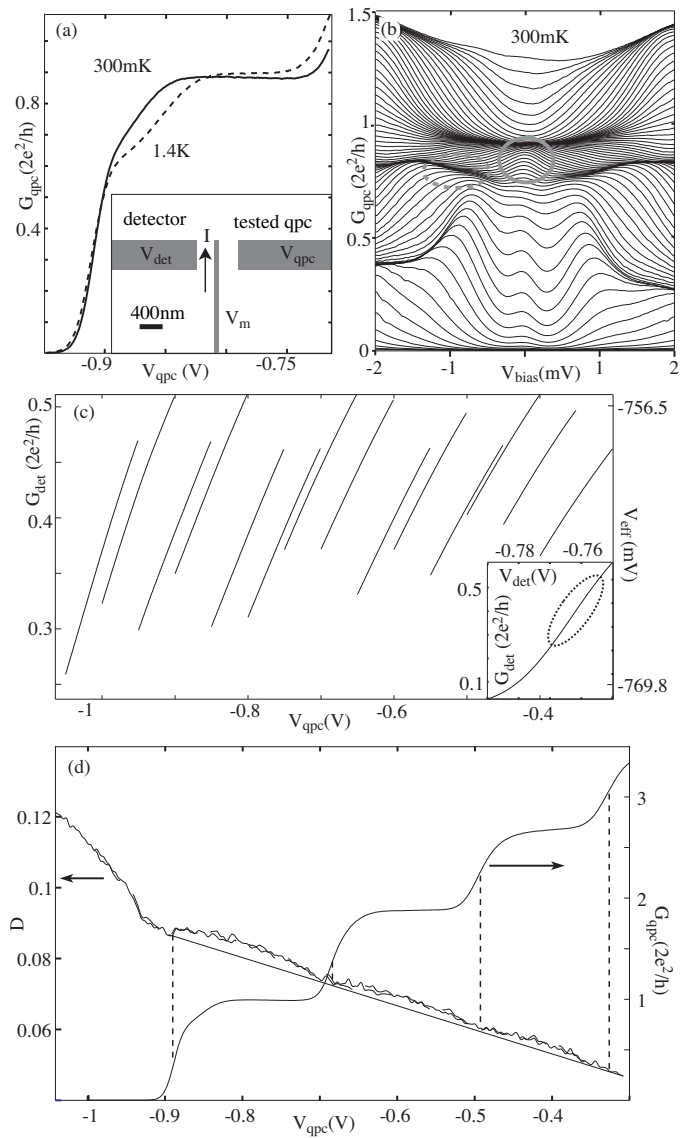}
         \caption[]{{\small Linear (a) and non-linear (b) conductance
         of the device sketched in inset of
         (a). Plateaux fall below \Go due to finite series resistance, not subtracted off in figures. Characteristic signatures \cite{Sara} of 0.7 structure are observed:
         high-bias plateau near $0.8(2e^2/h)$ (dashed ellipse) and
         zero-bias anomaly (solid ellipse). c) Inset: Conductance of
         the detector QPC, used to derive the mapping $V_{\rm
         eff}(V_{\rm qpc})$.  Optimal working range of the detector
         indicated by dotted ellipse. Main panel: 14 charge sensing
         measurements covering wide range of $V_{\rm qpc}$.  Left
         axis: detector conductance $G_{\rm det}$.  Right axis: $V_{\rm eff}$
         mapped from $V_{\rm qpc}$ as described in text.  d)
         Conductance trace of the \test (right axis) and derivative
         $D\equiv dV_{\rm eff}/dV_{\rm qpc}$ of the detector data
         (left axis) for identical gate settings. $D$ is a measure of
         screening of electric fields by the \test. Dashed lines mark
         steep rises in the conductance of the \test, coinciding with
         suppressions in $D$ which indicate enhanced screening.}}
     \label{cosweep}
  \end{center}
\end{figure}

To analyze the data more carefully, we perform two transformations.
First, we assign to each $V_{\rm qpc}$ an effective voltage, $V_{\rm
eff}$, that would produce the same change in the detector
conductance if it were applied to the detector gate.  This mapping
is based on the measured
response of the detector $G_{\rm det}(V_{\rm det})$, Fig.~1(c)
inset.
The result of this transformation is shown in Fig.~1(c) (right
axis). Second, the derivative $D\equiv dV_{\rm eff}/dV_{\rm qpc}$
eliminates the offsets between consecutive traces.  This quantity
describes the relative coupling of the detector to $V_{\rm qpc}$ and
$V_{\rm det}$, and thus provides a quantitative measure of the
screening of $V_{\rm qpc}$ by mobile charges in and around the QPC.

The derivatives of the 14 measurements shown in Fig.~1(c) are
plotted in Fig.~1(d). The overlapping ranges of each curve agree,
allowing us to extract a continuous $D(V_{\rm qpc})$. The QPC \G is
superimposed on the measurement of $D$ in Fig.~1(d). Three important
features are observed. First, $D$ increases steadily as $V_{\rm
qpc}$ becomes more negative. Second, the slope of $D$ becomes larger
immediately beyond \po of the \test ($V_{\rm qpc} \leq -0.95$~V).
Third, each steep rise in the QPC conductance is accompanied by a
slight dip in $D$, indicated with dashed lines in Fig.~1(d).

These features may be qualitatively understood as follows. The first
and second features are caused by the reduction in screening of
$V_{\rm qpc}$ by the 2DEG, as $V_{\rm qpc}$ is made more negative
and pushes the
 2DEG away from the QPC. This recession becomes more rapid after \po. The
explanation of the dips associated with the opening of each  mode is
less straightforward. Intuitively, the divergence of the 1D density
of states (1DOS) of the electrons in the QPC at the entry of each
additional subband should be accompanied by an enhancement of the
screening and a dip in $D$. Less obviously, enhanced screening at
\po could also result from the exchange interaction between
conduction electrons.

To understand the features in $D$ both qualitatively and
quantitatively, we first simulate the device numerically, using M.
Stopa's SETE code~\cite{Stopa}. The simulation calculates
self-consistently the effective potential and the density in the QPC
as a function of voltages on the electrostatic top gates, but does
not include quantum corrections associated with the 1D constriction
in the QPC \cite{simu}. Inputs to the simulation include the 2DEG
growth parameters, geometry and voltages of the three gates, and
temperature. The potential landscape of the device is calculated for
a range of settings of $V_{\rm qpc}$ and $V_{\rm det}$.

In Fig.~2(a) we plot $D^{\rm SETE}$ extracted from the simulated
data using a procedure analogous to that described for $D$ above.
The quantitative match between the simulated $D^{\rm SETE}$ and the
measured $D$ is striking. In the simulation, as in the measurement,
the kink in $D^{\rm SETE}$ occurs exactly at \po~\cite{pinchoff}.
This confirms the association of the change in slope with an
electrostatic effect of emptying the point contact's saddle
potential.

\begin{figure}[]
  \begin{center}
 \includegraphics[width=0.5\textwidth]{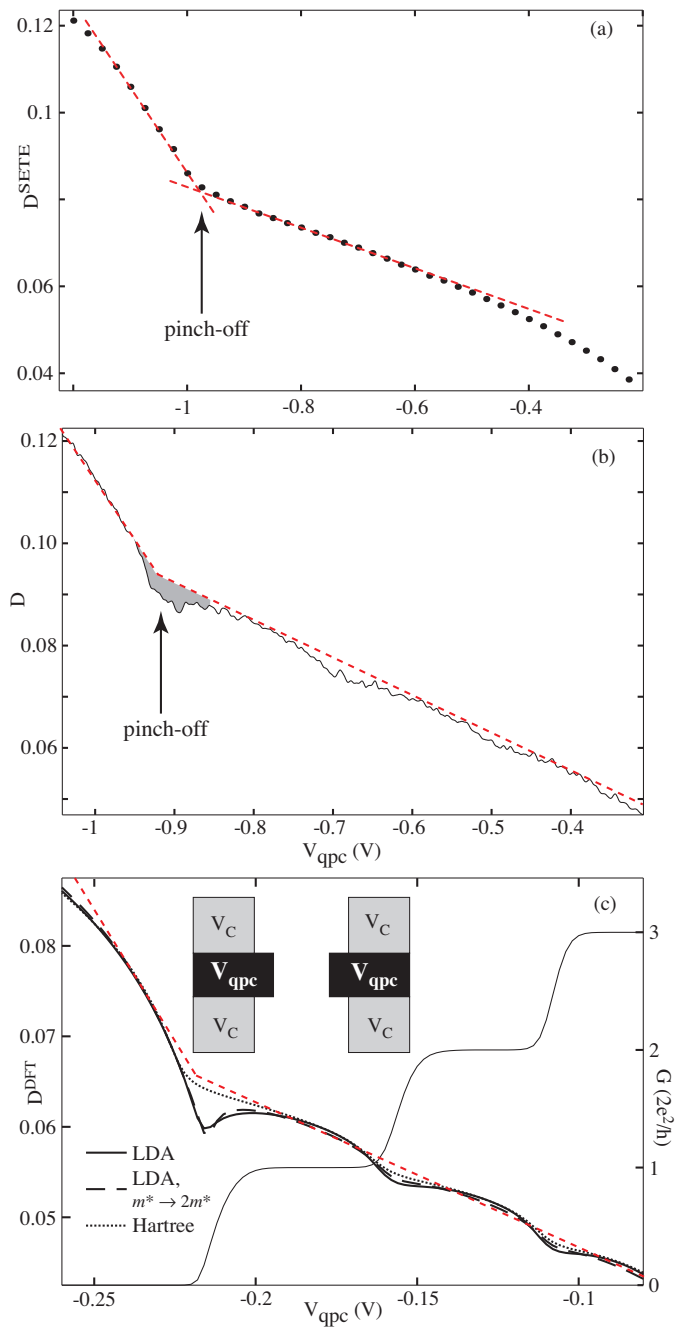}
  \caption[]{{\small a) $D^{\rm SETE}\equiv dV_{\rm eff}/dV_{\rm qpc}$
      extracted from numerical simulation of the device. The red dashed lines are
      guides to the eye, showing linearity of $D^{\rm SETE}$ around
      \po. No modulation can be observed in the simulation, which does
      not account for the one-dimensional nature of the
      QPC~\cite{simu}. b) The shaded area indicates a pronounced dip
      of $D$ around \po. No such feature is observed in the simulated
      $D^{\rm SETE}$. c) $D^{\rm DFT}$, as calculated from DFT
      simulation of a modified QPC (inset) with a detector
      200~nm beneath the plane of the QPC. Thick solid line: LDA with actual GaAs
      effective mass; dashed line: LDA with doubled mass; dotted line:
      Hartree approximation with actual mass. The thin solid line
      shows the conductance of the QPC in the LDA approximation
      (righthand scale). The dashed red line is again a guide to the eye, showing
      the change in slope of the detector signal at \po within all these simulations.
      The simulations dip below this guide line at the opening of each new mode. Within the Hartree approximation
      the dips are all roughly equal in size, but in LDA the dip at \po is substantially larger than the others,
      as observed in experiment.}}
  \end{center}
\end{figure}

Both the measured and the simulated derivatives are linear above and
below the kink at \po.
As seen in Fig.~2(a), the simulation is closely approximated by two
lines intersecting at the \po voltage.  Fig.~2(b) has analogous
lines to approximate the data above and below \po and hence to
emphasize the fact that the data drop below these guide lines at the
transitions between conductance plateaux. The absence of such dips
in the simulation, which includes only the classical electrostatic
effect of the gates on the 2DEG, supports the argument that the
modulations are caused by quantum mechanical effects such as a
nontrivial 1DOS or exchange interaction.

The data in Fig.~2(b) show a pronounced dip at \po. Understanding
the source of this dip and the smaller dips associated with opening
of successive subbands requires a quantum mechanical calculation.
Since such a calculation of the full three dimensional system is
very demanding, we perform a density functional theory (DFT)
simulation of a device in which the QPC is modeled by a constriction
in a long quantum wire that is wide enough to carry four
spin-degenerate modes~\cite{RejecNature,EPAPS}. In the inset of
Fig.~2(c), the quantum wire is defined by gates marked $V_{\rm c}$
and the constriction by gates marked $V_{\rm qpc}$.  We further
modify the dimensions and electron density to make the computation
feasible. The width and the length of the simulated QPC are 250~nm
and 200~nm, respectively, and the 2DEG is 70~nm below the surface.
The donor density is $10^{11}$~cm$^{-2}$. The screened potential
$V_{\rm det2}$ is detected 200~nm beneath the center of the QPC. We
simulate the device for a range of $V_{\rm qpc}$ such that the QPC
has from zero to three open subbands -- the thin solid line in
Fig.~2(c) shows the conductance. Screening of $V_{\rm qpc}$ by the
QPC affects the value of $D^{\rm DFT}\equiv \mathrm{d}V_{\rm
det2}/\mathrm{d}V_{\rm qpc}$. We use the local density approximation
(LDA) for the exchange-correlation energy of the electrons in the
2DEG, using the known effective mass and dielectric constant for
GaAs. Because of the modified geometry and the different electron
density the DFT simulation cannot reproduce our measured $D$
quantitatively. However the simulation clearly shows a dip whenever
a new subband opens. In agreement with the measured data, the dip at
the opening of the first subband is more pronounced than those for
the second and third subbands. To estimate the importance of
exchange and correlation we eliminate them by making the Hartree
approximation (dotted curve)~\cite {Hartree}. The dips are much
weaker in this approximation (they disappear entirely in the
Thomas-Fermi approximation) and all dips are of similar size,
demonstrating that the larger dip at \po in the LDA calculation is
dominated by exchange-correlation effects. This is not surprising,
since the exchange-correlation contribution to total energy at low
carrier density is larger than the kinetic contribution: for
quasi-1D systems, these two contributions go as $n^{3/2}$ and $n^3$,
respectively, where $n$ is the linear density of electrons. The
prominence of the dip at \po relative to those for higher modes in
the experiment (Fig.~2(b)) is thus evidence of the importance of
exchange-correlation effects to charge distribution in a QPC. For
higher modes, kinetic energy plays an important role: lowering
kinetic energy by doubling effective mass (dashed curve, Fig.~2(c))
reduces dip area by 40\%.

\begin{figure}[]
\begin{center}
\includegraphics[width=0.9\columnwidth,keepaspectratio]{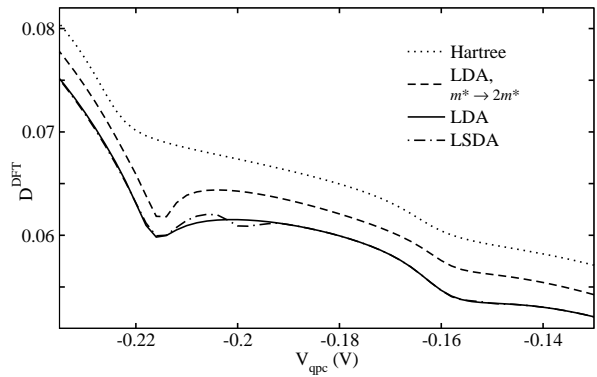}
\caption[]{{\small Curves from Fig.~2(c) expanded and shifted
    vertically to highlight subtle differences
    among the
    approximations at the opening of the
    first and second subbands. The LSDA curve with an additional dip
   associated with spin rearrangement is also shown, unshifted with
   respect to the LDA curve.
   }}
\end{center}
\end{figure}

To see whether spin rearrangement within the QPC should be
detectable by our compressibility measurements, we have also
performed a DFT simulation using the local spin density
approximations (LSDA) in which the density of electrons with spin up
is not restricted to be the same as that for spin down. In this case
the simulated detector signal has an additional small dip (Fig.~3,
dot-dashed), marking the subtle rearrangement of electron density
associated with the formation of a quasibound state at the center of
the QPC. This smaller dip would be a signature of the formation of a
magnetic moment responsible for the Kondo effect. Only one broad dip
is discernible in the experimental data, precluding a clear
experimental statement about spin rearrangement.

In conclusion, we have used a charge-sensitive detector to measure
the charge rearrangement in a QPC as it passes through \po. The
broad features in the detector signal closely match the predictions
of a classical electrostatic simulation of the device. However, the
charge redistribution at the entry of 1D subbands creates a series
of dips in the detector signal. DFT calculations provide a good
qualitative match to measurement, and indicate that the first dip is
dominated by exchange interaction between electrons. The other dips
are associated with the divergence of the density of states at the
opening of each subband. Similar DFT calculations which further
allow for the formation of local (spin-degenerate) magnetic
polarization serve as the basis for the local moment Kondo scenario
for 0.7 structure~\cite{Sara}. Our measurements are consistent with
both LDA and these LSDA calculations, since in the calculations the
spin rearrangement is accompanied by only a slight charge
rearrangement.

We thank the NNIN Computation Project for SETE, and L. P.
Kouwenhoven and L. H. Willems van Beveren for computer access. Work
at Stanford was supported by contracts from the US AFOSR
(F49620-02-1-0383) and the US ONR (YIP, N00014-01-1-0569). Work at
BGU was supported by the BSF. SL acknowledges a SNF Fellowship, DGG
support from the Sloan Foundation and Research Corporation, and TR
support from the Kreitman Foundation. During the review process, we
became aware of related work on carbon nanotubes~\cite{Ilani2006}.

\newpage
\providecommand{\boldsymbol}[1]{\mbox{\boldmath $#1$}}

\renewcommand{\thefigure}{S\arabic{figure}}
\setcounter{figure}{0}

\noindent \begin{center}{\Large Supplementary Information: Results
of the DFT calculation with the potential detected in the 2DEG
plane}\par\end{center}{\Large
\par}

In the DFT calculation we unfortunately cannot simulate a device in
which both the measured QPC and the detector QPC are present. In
order to perform such a calculation we would have to put both QPCs
inside a very wide quantum wire. The number of modes such a wire
would have to support would be too large for the calculation to be
feasible. In the single QPC geometry we used in our DFT calculation
we could have measured the potential in the 2DEG plane next to the
QPC instead of 200 nm below the 2DEG plane as we did. It turns out
that the features observed in the experiment and in our DFT
calculation can also be observed if we simulate measuring the
potential in the 2DEG plane (see Figure S1.) In Figure S2, the
derivative $D$ was calculated from the potential measured in the
2DEG plane 300 nm from the center of the QPC. As in the calculation
presented in the paper, there is a large dip at pinch-off and a
series of weaker dips corresponding to higher transverse modes being
occupied in the QPC. Observation of these features in two different
geometries supports our contention that they are qualitatively
linked to the features we observe in the experiment. However,
because the experimental geometry is different from that of the two
calculations we cannot claim that the calculations should
quantitatively account for the magnitudes of the dips in the real
device.
\begin{figure}[H]
\begin{centering}\includegraphics[width=0.9\columnwidth]{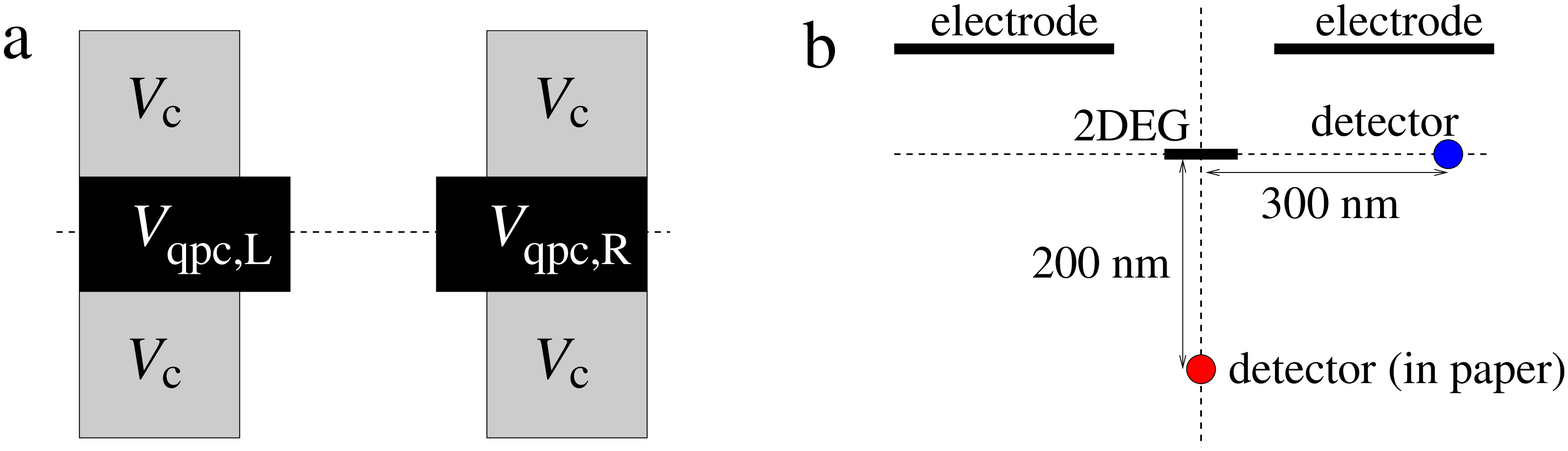}\par\end{centering}
\caption{a) The electrodes of the modified QPC used in the DFT
calculation.  b) The cross section through the center of the QPC
(dashed line in panel a). The red dot 200 nm below the 2DEG is the
position of the detector in the calculation presented in the paper.
The derivative $D^{DFT}$ calculated from the potential in the 2DEG
plane 300 nm from the center of the QPC (blue dot) is shown in the
next figure. In the calculation presented in the paper the voltages
on both gate electrodes were changed together ($V_{\mathrm{qpc, L}}
= V_{\mathrm{qpc, R}}$), while in the present calculation
$V_{\mathrm{qpc, R}}$ was fixed and we only varied $V_{\mathrm{qpc,
L}}$.  }
\end{figure}

\begin{figure}[H]
\begin{centering}\includegraphics[width=0.9\columnwidth]{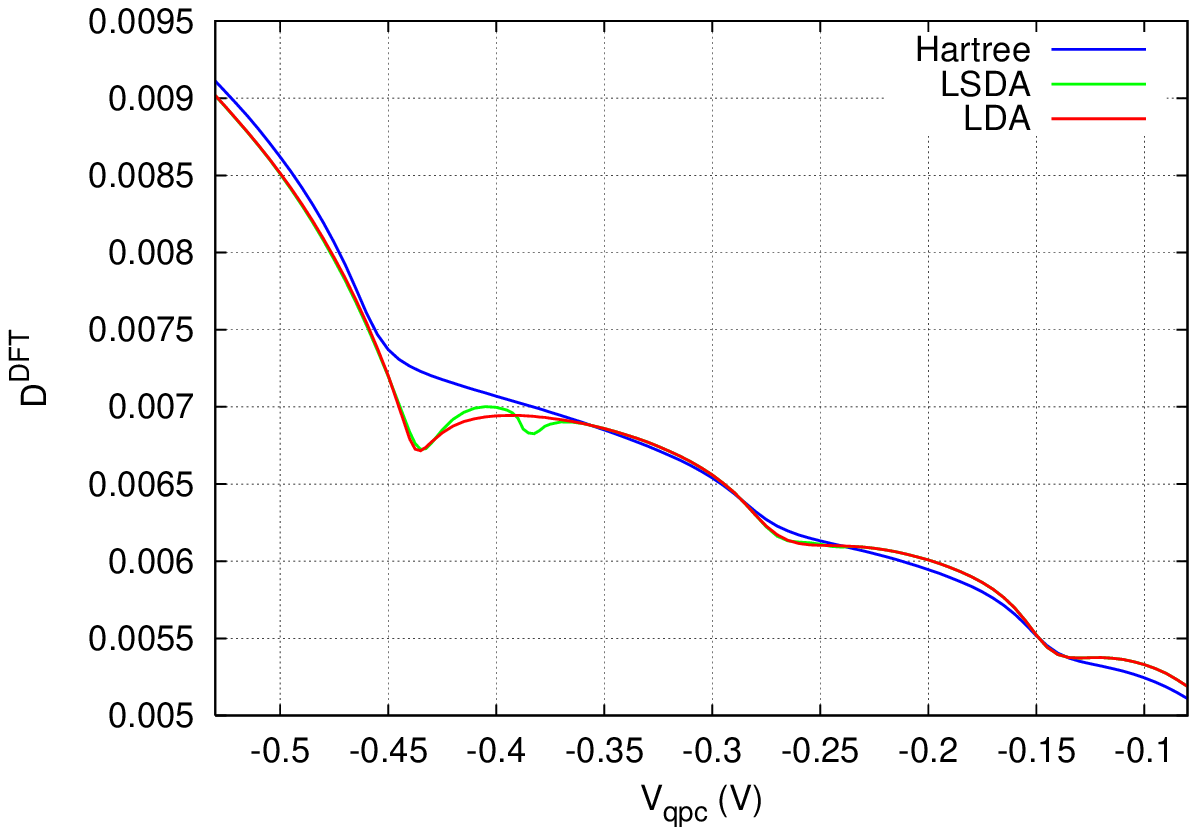}\par\end{centering}

\caption{The derivative $D^{DFT}$ calculated from the potential off
to the side in the 2DEG plane (detector location is shown as a blue
dot in Fig. S1b). The curves for three different approximations are
shown: DFT within local density approximation (red), DFT within
local spin-density approximation (green) and the Hartree
approximation (blue). The curves show very similar behavior to that
of the calculation presented in the paper (Fig. 2c and Fig. 3),
showing that the detailed geometry is not important in determining
the qualitative features of the compressibility measurement, as long
as the system under study lies between the gate that is varied and
the detector. The magnitude of $D^{DFT}$, however, is smaller here
because the right gate electrode effectively screens the potential.
In the experiment, the electrode separating the two QPCs is very
thin and is thus less effective in screening the potential. }
\end{figure}

\noindent \begin{center}{\Large Adjustable
parameters}\par\end{center}{\Large
\par}
We did not adjust parameters to try to achieve the best possible fit
to the data. Rather, we chose values for parameters as close as
possible to the values in the real device while maintaining
computational feasibility.

Values for most of the important parameters for the DFT calculation
are given in the manuscript. Other relevant parameters are the
lithographic width of the quantum wire away from the QPC (300 nm),
the voltage on the quantum wire electrodes ($V_c = -0.08 V$) and the
position of the donor layer (20 nm below the surface). The details
of our numerical approach are presented in a separate
publication~\cite{RejecNature}.

\newpage
\noindent \begin{center}{\Large Can spin rearrangement be detected
with our method?}\par\end{center}{\Large
\par}
The additional dip near pinch-off in the local spin-density
approximation (LSDA) calculation is related to formation of a
spin-1/2 magnetic moment in the QPC. Somewhere between pinch-off and
the transition to the first conductance plateau, a quasibound state
forms at the center of the QPC just below the Fermi energy. The
quasibound state can be occupied with either a spin-up or a
spin-down electron, i.e. the QPC acts as a spin-1/2 magnetic moment.
The formation of the magnetic moment is accompanied by a slight
rearrangement of the electron density within the QPC, which shows up
as an additional dip in $D$.

As alluded to above, some of the present authors have performed an
extensive study of the conditions under which such a magnetic moment
forms in the QPC~\cite{RejecNature}. We observed this feature only
in the DFT calculation within the local spin-density approximation.
In the ordinary local density approximation the two spin-densities
are restricted to be the same and thus spin-polarization in the QPC
is not possible, while in the Hartree approximation the exchange
interaction, which is responsible for spin-polarization, is absent.
We observed the formation of the magnetic moment for various
geometries of the QPC (rectangular shape of QPC electrodes or more
adiabatic, triangular shaped electrodes) and for a wide range of QPC
widths and 2DEG densities. The parameter which affects the formation
of the quasibound state in the most important way is the length of
the QPC. For a substantial range of QPC lengths the situation is the
same as presented in this paper. However, if the QPC is very short
there is not enough space for the quasibound state to form in it. If
the QPC is very long an antiferromagnetically ordered spin chain
forms instead of a single spin-1/2 magnetic moment.

Our present experiment cannot resolve the extra dip described by the
LSDA calculation, but future, more sensitive measurements may be
able to detect it.


\begin{thebibliography}{99}
\bibitem{simu} The simulation solves Schr\"{o}dinger and Poisson equations in the growth direction
at each point on a grid in the plane of the 2DEG. Then it calculates
the density self-consistently using a Thomas-Fermi approximation.
\bibitem{pinchoff} Pinch-off in the simulation is defined as the point where the conduction band edge
is equal to the Fermi energy exactly at the saddle point.
\bibitem{Hartree} The exclusion of exchange and correlation affects
the density far away from the QPC. To partly compensate for this
effect we shifted the Hartree curve so that it matches the DFT
result far beyond pinch-off.
\end{thebibliography}
\end{document}